\documentclass[prd,aps,showpacs,twocolumn,preprintnumbers,nofootinbib,amsmath,amssymb,superscriptaddress,longbibliography]{revtex4-2}
\allowdisplaybreaks

\usepackage{systeme}
\usepackage{mathtools}
\usepackage[inline]{enumitem}
\usepackage{graphicx}

\usepackage[T1]{fontenc}
\usepackage[english]{babel}
\usepackage{xcolor}
\usepackage{braket}
\usepackage{nicefrac}
\usepackage{bm}
\usepackage{bbm}
\usepackage{braket}
\usepackage{bbold}    %for unit matrix symbol
\usepackage{slashed}
\usepackage[normalem]{ulem}
\usepackage{array}
\newcolumntype{C}{>{$}c<{$}}

\newcommand{\beqn}{\begin{eqnarray}}
\newcommand{\eeqn}{\end{eqnarray}}
\newcommand{\beqs}{\begin{subequations}}
\newcommand{\eeqs}{\end{subequations}\\[-2mm]\noindent}
\newcommand{\cO}{{\mathcal O}}
\newcommand{\cC}{{\mathcal C}}
\newcommand{\eq}[1]{(\ref{#1})}

\newcommand{\bs}{\boldsymbol}
\newcommand{\avr}[1]{{\left\langle #1 \right\rangle}}
\allowdisplaybreaks
\definecolor{brickred}{rgb}{0.8, 0.25, 0.33}
\definecolor{macouleur}{RGB}{105,150,150}

\usepackage{color}
\definecolor{purple}{rgb}{0.8,0,0.6}

\begin{document}

\bibliographystyle{apsrev4-1}

\title{Inhomogeneity of rotating gluon plasma and Tolman-Ehrenfest law in imaginary time:\\ lattice results for fast imaginary rotation}

\author{M. N. Chernodub}
\affiliation{Institut Denis Poisson UMR 7013, Universit\'e de Tours, 37200 Tours, France}

\author{V. A. Goy}
\affiliation{Pacific Quantum Center, Far Eastern Federal University, 690950 Vladivostok, Russia}

\author{A. V. Molochkov}
\affiliation{Pacific Quantum Center, Far Eastern Federal University, 690950 Vladivostok, Russia}

\begin{abstract}
We present the results of first-principle numerical simulations of Euclidean SU(3) Yang-Mills plasma rotating with a high imaginary angular frequency. The rigid Euclidean rotation is introduced via  ``rotwisted'' boundary conditions along imaginary time direction. The Polyakov loop in the co-rotating Euclidean reference frame shows the emergence of a spatially inhomogeneous confining-deconfining phase through a broad crossover transition. A continuation of our numerical results to Minkowski spacetime suggests that the gluon plasma, rotating at real angular frequencies, produces a new inhomogeneous phase possessing the confining phase near the rotation axis and the deconfinement phase in the outer regions. The inhomogeneous phase structure has a purely kinematic origin, rooted in the Tolman-Ehrenfest effect in a rotating medium. We also derive the Euclidean version of the Tolman-Ehrenfest law in imaginary time formalism and discuss two definitions of temperature at imaginary Euclidean rotation.
\end{abstract}

\date{\today}

\maketitle

\section{Introduction}

Noncentral collisions of relativistic heavy ions produce quark-gluon plasma with extraordinarily high vorticity. The experimental results of the STAR collaboration at RHIC indicate that the vorticity of the rotating plasma reaches the values $\Omega \approx (9 \pm 1)\times 10^{21} \, \mathrm{s}^{-1} \sim 0.03\, \mathrm{fm}^{-1} c \sim 7 \, \mathrm{MeV}$~\cite{STAR:2017ckg} while the theoretical analysis predicts that the plasma can rotate even faster depending on the initial parameters of the collisions~\cite{Deng:2016gyh,Jiang:2016woz}.

One expects that sufficiently fast rotation influences the local properties of quark-gluon plasma leading to a series of spin polarization effects~\cite{Becattini:2020ngo,Huang:2020dtn} that also allow us to interpret the experimental response of rotating plasma fireball in terms of its local vortical structure. Various theoretical arguments suggest that vorticity also affects the thermodynamic characteristics of the quark-gluon plasma. Consequently, the rotation was proposed to modify the phase diagram by shifting the existing transition line which separates the hadronic and plasma phases~\cite{Chen:2015hfc,Jiang:2016wvv,Chernodub:2016kxh,Chernodub:2017ref,Wang:2018sur,Zhang:2020hha,Sadooghi:2021upd,Braguta:2020biu,Braguta:2021jgn} and also by introducing a new inhomogeneous phase characterized by spatial phase separation due to rotation~\cite{Chernodub:2020qah}. 

Theoretical approaches often consider a rigid rotation of the quark-gluon plasma. Although the rigid nature of the rotation drastically simplifies the analytical treatment of the system~\cite{Ambrus:2014uqa,Ambrus:2015lfr}, the global consensus on the thermodynamic properties of rotating quark-gluon plasma is still lacking. Our work considers a globally rotating gluon plasma using analytical and first-principle numerical methods. 

The finite-temperature QCD phase transition is accompanied by deconfinement of color and chiral symmetry restoration. There is a general agreement in the community that the uniform rotation reduces the temperature of the chiral phase transition~\cite{Chen:2015hfc, Jiang:2016wvv, Chernodub:2016kxh, Chernodub:2017ref, Wang:2018sur, Zhang:2020hha, Sadooghi:2021upd} implying that the global rotation should restore the chiral symmetry at lower critical temperatures than it happens in non-rotating quark-gluon plasmas. The mechanism behind this phenomenon is a generalization of the Barnett effect~\cite{Barnett:1915} found in 1915: the rotation tends to align the spins of quarks and anti-quarks along the rotation axis, thus suppressing the scalar pairing and, therefore, lowering the scalar fermionic condensate~\cite{Jiang:2016wvv}.

Contrary to the chiral properties of quark-gluon plasma, the first numerical simulation of pure gluon plasma has revealed that the bulk critical temperature of the deconfining phase transition increases with the increase of the global rotation frequency~\cite{Braguta:2020biu}. This conclusion, achieved in lattice SU(3) Yang-Mills theory, has been confirmed in the subsequent study~\cite{Braguta:2021jgn} where independence of the bulk effect on the type of spatial boundary conditions has also been reported. 

However, two independent theoretical approaches to the same problem, a holographic technique of Ref.~\cite{Chen:2020ath} and the effective model of the hadron resonance gas in Ref.~\cite{Fujimoto:2021xix} give the opposite outcome implying that the temperature of the deconfinement decreases as the vorticity of the plasma raises. While the second scenario complies with the chiral properties of the rotating plasma~\cite{Chen:2015hfc, Jiang:2016wvv, Chernodub:2016kxh, Chernodub:2017ref, Wang:2018sur, Zhang:2020hha,Yadav:2022qcl} (in the assumption that the chiral and deconfining transitions happen simultaneously under rotation), they contradict the lattice data of Refs.~\cite{Braguta:2020biu, Braguta:2021jgn}.

A third scenario of the rotational effect on the phase diagram has been put forward in Ref.~\cite{Chernodub:2020qah}, arguing that rotation leads to a qualitative change of the QCD phase diagram leading to a new mixed confining-deconfining phase. Consequently, the deconfining phase transition, inherent to the non-rotating plasma, should split at nonvanishing angular frequency, $\Omega \neq 0$, into two deconfining transitions: the first transition at $T = T_{c1}(\Omega)$ separates the pure confinement phase and the new mixed inhomogeneous phase, while the second transition at $T = T_{c2}(\Omega) > T_{c1}(\Omega)$ separates the mixed and pure deconfinement phases at higher temperatures. This scenario is different from the existing observations that the relativistic rotation can lead to inhomogeneities in a single phase (for example, for the chiral condensate in the low-temperature phase~\cite{Chen:2022mhf}). 

In the Euclidean imaginary time formalism, accessible to lattice simulations, the real angular momentum $\Omega$ becomes the imaginary quantity $\Omega_I = - i \Omega$ similarly to the baryon chemical potential~\cite{Yamamoto:2013zwa,Braguta:2020biu,Chernodub:2020qah,Braguta:2021jgn,Chen:2022smf,Chernodub:2022wsw}. The Euclidean Yang-Mills theory at imaginary rotation is interesting by itself~\cite{Chen:2022smf}, especially for the discussion of the justification of analytical continuation to real rotation used in numerical lattice simulations~\cite{Braguta:2020biu,Braguta:2021jgn} and analytical works~\cite{Chernodub:2020qah}. Moreover, the Euclidean action in the curved metric corresponding to the imaginary angular frequency has no causality-related singularity~\cite{Yamamoto:2013zwa,Braguta:2020biu,Braguta:2021jgn} so that the Euclidean theory with imaginary rotation has a well-defined thermodynamic limit~\cite{Chen:2022smf}. There is also an interesting implementation for the topology of gauge fields: the Yang-Mills instanton under the imaginary rotation de-localizes over constituents that carry fractional topological charge. It becomes the axially symmetric ``circulon'' solution in a high-temperature limit~\cite{Chernodub:2022wsw} (see also earlier discussion of the discrete rotational map for calorons~\cite{Cork:2017hnj}).

The structure of this paper is as follows. Section~\ref {sec_real_vs_imaginary} discusses theoretical aspects of real and imaginary rotations, including the Tolman-Ehrenfest effect in Minkowski and Euclidean spacetimes, the phase diagram of Yang-Mills theory under real and imaginary rotation, and the validity of the analytical continuation under the Wick transformation. Section~\ref {sec_lattice} presents the first-principle numerical investigation of the spatial structure of the rapidly rotating quark-gluon plasma, which allows us to clarify some of the mentioned properties and also find new puzzles. Finally, the last section is devoted to conclusions and discussions.

\section{Real vs Imaginary Rotations}
\label{sec_real_vs_imaginary}

\subsection{Rotation and the Tolman-Ehrenfest effect}

The Tolman-Ehrenfest (TE) effect implies that the local temperature $T = T({\bs x})$ of a system residing in a global thermal equilibrium in a time-independent gravitational field is an inhomogeneous quantity~\cite{Tolman:1930zza, Tolman:1930ona}:
\beqn
T({\bs x}) \sqrt{g_{00}(\bs x)} = T_0\,, 
\label{eq_TE}
\eeqn
where $g_{00}$ is the component of the metric tensor and $T_0$ is a reference temperature.

Let us apply the TE law~\eq{eq_TE} to a body that rotates rigidly with the constant angular frequency $\Omega$ around the $z$ axis. The local temperature is defined in the co-rotating reference system in which the rotating body appears static. In cylindrical coordinates, $x^\mu = (\rho, \varphi, z, t)$, the co-rotating reference frame is given by a curvilinear metric with the line element
\beqn
d s^2 = g_{\mu\nu} d x^\mu dx^\nu & = & \left( 1 - \Omega^2 \rho^2 \right) d t^2 - 2 \Omega \rho^2 d t d \varphi \nonumber \\
& & - d \rho^2 - \rho^2 d \varphi^2 - d z^2\,.
\label{eq_d_s}
\eeqn
Reading off the relevant metric element from Eq.~\eq{eq_d_s}, $g_{00} = 1 - \Omega^2 \rho^2$, one finds that the local temperature~\eq{eq_TE} is an increasing function of the radial distance~$\rho$:
\beqn
T_{\mathrm{TE}}(\rho) = \frac{T_0}{\sqrt{1 - \Omega^2 \rho^2}}\,,
\label{eq_T_rho}
\eeqn
where the subscript ``TE'' stands for the Tolman-Ehrenfest law. The quantity $T_0$ in Eq.~\eq{eq_T_rho} corresponds to the local temperature on the axis of rotation $T_0 \equiv T_{\mathrm{TE}}(\rho = 0)$. The causality requirement enforces the bound $\rho |\Omega| < 1$, which defines the light cylinder of a rotating system (a rotational analog of the light cone). If this limit is violated, the first term in Eq.~\eq{eq_T_rho} turns negative, and the line element becomes imaginary. 

The emergence of the new inhomogeneous phase in QCD (and in Yang-Mills theory) becomes apparent after a lengthy calculation in an analytically solvable confining model~\cite{Chernodub:2020qah}. However, the physical origin of this inhomogeneous phase has a simple kinematic reason rooted in the simple TE relation~\eq{eq_T_rho} stating that the temperature of the rotating system is higher in the peripheral regions (at the largest $\rho \neq 0$) as compared to its center ($\rho = 0$). 

Consider a system of a cylindrical geometry of the radius $R$ rotating with a constant angular velocity $\Omega$ residing in thermal equilibrium. At vanishing temperature at the center, $T_0 = 0$, the global temperature of the cylinder is zero, so that $T(\rho) = 0$ for all $\rho$. Now, let us gradually increase the temperature $T_0$ at the center of the sample. As the peripheral layers of the system are always hotter than the interior, according to Eq.~\eq{eq_T_rho}, the deconfining temperature $T = T_{c1}$ will be achieved first at the boundary $\rho = R$. The system will enter the mixed deconfining phase (with confined interior and deconfined exterior) above the first critical temperature~\cite{Chernodub:2020qah}:
\beqn
T_{c1} = T_{c,\infty} \sqrt{1 - \Omega^2 R^2}\,.
\label{eq_Tc_1}
\eeqn
Here $T = T_{c,\infty}$ is the deconfining transition in the thermodynamic limit of a non-rotating system. When this temperature is reached, the whole system becomes deconfined. The global deconfinement in the whole space appears at the second critical point:
\beqn
T_{c2} = T_{c,\infty}\,.
\label{eq_Tc_2}
\eeqn
Thus, we have a confining and deconfining phases below $T_{c1}$ and above $T_{c2}$, respectively, with the new inhomogeneous phase emerging at the intermediate range of temperatures, $T_{c1} < T < T_{c2}$. 

\begin{figure}[ht]
\centering
  \includegraphics[width=1\linewidth]{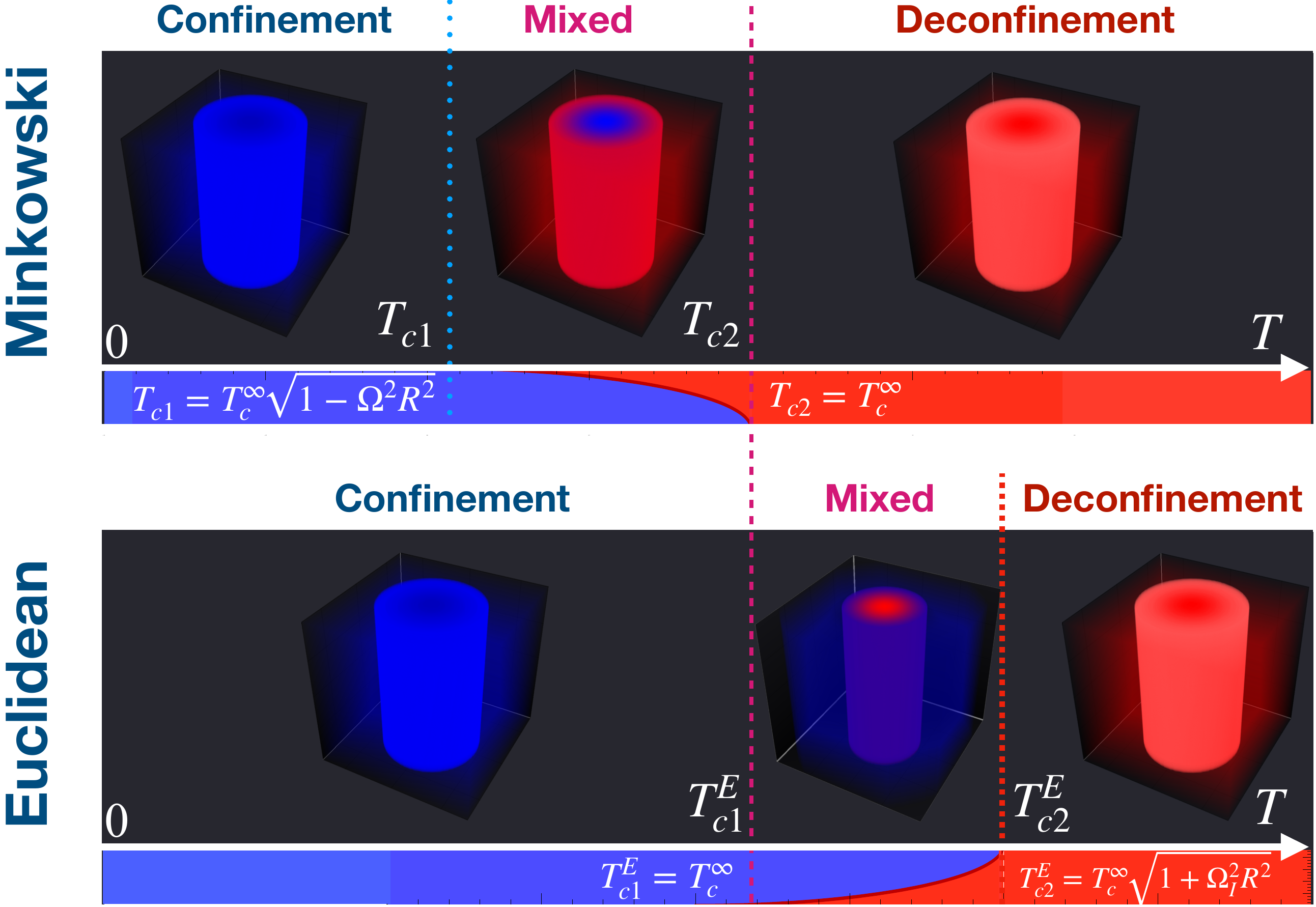}
  \caption{Suggested phase structure of rotating (quark-)gluon plasma at finite temperature $T$: (a) Effect of real rotation in Minkowski spacetime: the phase diagram of gluon plasma in a cylinder of the radius $R$ which rotates rigidly with the angular frequency~$\Omega$; (b) Effect of imaginary rotation in Euclidean space: the phase diagram of gluon plasma in an infinite volume which rotates rigidly with the imaginary angular frequency $\Omega_I$. The upper panel is adopted from Ref.~\cite{Chernodub:2020qah}.}
  \label{fig_phaselines}
\end{figure}

\begin{figure}[ht]
\centering
  \includegraphics[width=1\linewidth]{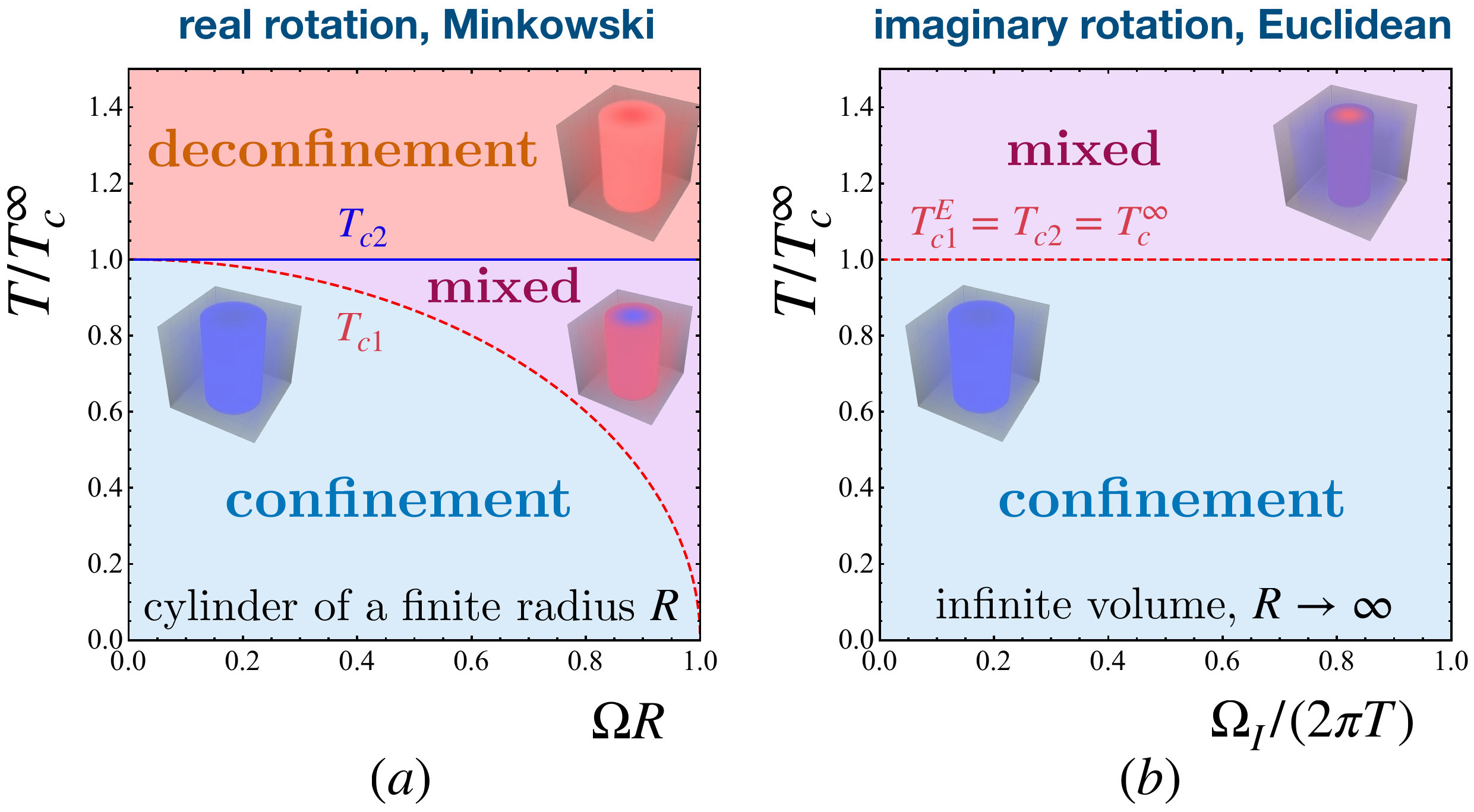}
  \caption{The phase diagram (a) in $(T,\Omega)$ plane for real rotation in Minkowski space-time; (b) in $(T,\Omega_I)$ plane for imaginary rotation in Euclidean, imaginary time formalism. The insets illustrate the spatial phase structure following Fig.~\ref{fig_phaselines}.}
  \label{fig_phase_diagram}
\end{figure}

The phase structure of gluon plasma of cylindrical geometry rotating with a fixed frequency $\Omega$ is shown, as a function of temperature $T$, in the upper panel of Fig.~\ref{fig_phaselines}. The same phase diagram, now depicted in the $(T,\Omega)$ plane, is illustrated in Fig.~\ref{fig_phase_diagram}(a). 

\subsection{Inverse hadronization effect}

The three-phase structure of the rotating gas is the consequence of the TE law~\eq{eq_TE} which has a kinematic origin related to the simple property that thermal wavelength gets red-shifted (or blue-shifted) as hot matter traverses the static gravitational field. The TE law also leads to a counter-intuitive ``inverse hadronization effect'' implying that as the global temperature (dictated by $T_0$) of the rotating quark-gluon plasma decreases, the hadronization occurs first at the axis of rotation, followed by the hadronization at the boundary~\cite{Chernodub:2020qah}. The inverse hadronization effect contradicts our daily experience, which tells us that the cooling (``hadronization'') of a hot system being in contact with a colder environment first starts from its boundary and not from its interior. Of course, in the first example, there is no contact with the external environment where the cooling in the center of rotation appears as the relativistic kinematic phenomenon caused by the TE effect.

\subsection{Analytical continuation}

The numerical simulations in thermodynamic equilibrium are performed in the Euclidean spacetime after the Wick transformation\footnote{This transformation is usually called ``the Wick rotation''. To avoid confusion, we will call it ``transformation'' instead of ``rotation'', since in our article, we are also considering real- and imaginary-space rotations.} from Minkowski spacetime, 
\beqn
t \to - i \tau\,.
\label{eq_Wick}
\eeqn
Here, the arrow, ``$\to$'', means ``identified with''. Under the Wick transformation, the angular frequency becomes an imaginary quantity similar to, for example, a baryon chemical potential. Indeed, the angular frequency corresponds to an angle at which the system turns per a unit of time. Under the identification~\eq{eq_Wick}, the time variable becomes an imaginary quantity (while the angle always stays real), suggesting that it is natural to consider the imaginary frequency:
\beqn
\Omega_I = - i \Omega\,.
\label{eq_Omega_I}
\eeqn

The first-principle lattice simulations of rotating systems are performed at the imaginary angular frequency $\Omega_I$~\cite{Yamamoto:2013zwa}. To obtain the results for real physical systems rotating at real-valued frequencies $\Omega$, an analytical continuation of the lattice results to the real frequency domain is usually done using the simple identification~\cite{Braguta:2020biu,Braguta:2021jgn}:
\beqn
\Omega_I^2 \leftrightarrow - \Omega^2\,.
\label{eq_Omega_I_2}
\eeqn

The analytical continuation~\eq{eq_Omega_I_2} has certain subtle features that could make this procedure questionable~\cite{Chen:2022smf}. This property could be the reason of the inconsistency of the lattice results~\cite{Braguta:2020biu,Braguta:2021jgn}, obtained via the analytical continuation~\eq{eq_Omega_I_2}, with the predictions of the analytical models~\cite{Chen:2020ath,Fujimoto:2021xix,Yadav:2022qcl} and \cite{Chernodub:2020qah}. The properties of real and imaginary rotations are considered in detail for a free scalar field in Ref.~\cite{Chernodub:2020qah}. While we also briefly discuss the analytical continuation from imaginary to real angular frequencies, our paper is mostly devoted to the Euclidean system rotating with an imaginary frequency $\Omega_I$.

\subsection{Imaginary rotation on Euclidean space}

In lattice gauge theory in Euclidean spacetime, the choice of the imaginary $\Omega_I$ is favored over the real $\Omega$ because in the latter case, the action becomes imaginary (and, thus, inappropriate for Monte Carlo algorithms) while the former choice guarantees its real-valuedness~\cite{Yamamoto:2013zwa}. The imaginary rotation $\Omega_I$ in lattice simulations can be introduced in two ways. 

First, one can identify the lattice action in the curved Euclidean spacetime with the following Euclidean distance element~\cite{Yamamoto:2013zwa,Braguta:2020biu,Braguta:2021jgn}:
\beqn
d s^2_E = g^E_{\mu\nu} d x^\mu dx^\nu & = & \left( 1 + \Omega^2_I \rho^2 \right) d \tau^2 + 2 \Omega_I \rho^2 d \tau d \varphi \nonumber \\
& & + d \rho^2 + \rho^2 d \varphi^2 + d z^2\,.
\label{eq_d_s_E}
\eeqn
Equation~\eq{eq_d_s_E} follows from its rigidly rotating Minkowski counterpart~\eq{eq_d_s} by applying the Wick transformation to the time variable~\eq{eq_Wick}, passing to the imaginary frequency~\eq{eq_Omega_I}, and identifying the line elements as follows: $d s^2 \to - d s^2_E$. 

Second, in the imaginary time formalism, the imaginary rotation can be understood as a geometrical rotation in the Euclidean spacetime~\cite{Chernodub:2020qah,Chen:2022smf}. As the imaginary time $\tau$ increases, the system uniformly rotates with the imaginary angular frequency $\Omega_I$ around a (``spatial'') axis that is normal to the time direction. 

\begin{figure}[ht]
\centering
  \includegraphics[width=0.85\linewidth]{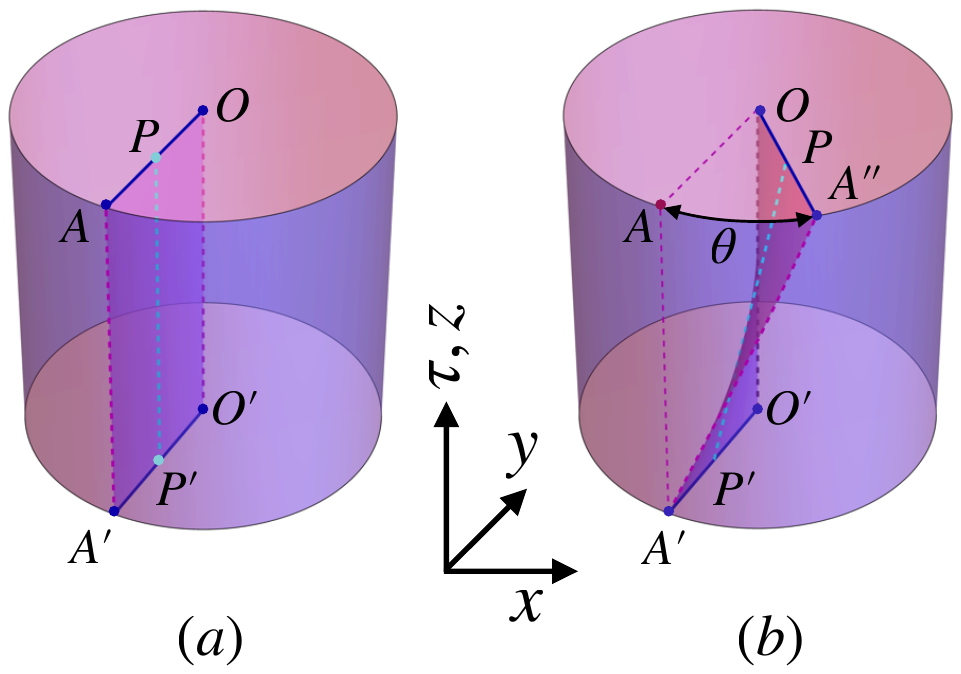}
  \caption{(a) The periodic boundary conditions for a non-rotating system: Illustration of a finite-temperature non-rotating ($\Omega_I = 0$) system with the standard boundary conditions along the compactified imaginary time $\tau$: the points $A$ and $A'$, $P$ and $P'$, as well as, respectively, $O$ and $O'$, are identified pairwise as the imaginary time advances for the full period, $\tau \to \tau + \beta$. (b) Rotwisted boundary conditions with the same identifications of the points at the imaginary rotation~\eq{eq_phi_rotation_cyl} at $\Omega_I \neq 0$. The spatial three-dimensional space rotates at the angle $\theta$, given in Eq.~\eq{eq_theta}, about the axis $z$ before the the identification of the $\tau = 0$ and $\tau = \beta$ time slices.}
  \label{fig_cylinders}
\end{figure}

At finite temperature $T$, the Euclidean direction is compactified to the circle $S^1_\tau$ of the length $\beta = 1/T$. In the absence of rotation, the boundary conditions in the circle $S^1_\tau$ are periodic for the bosonic fields, $\phi({\bs x},\tau) = \phi({\bs x},\tau + \beta)$ and anti-periodic for fermionic fields, $\psi({\bs x},\tau) = - \psi({\bs x},\tau + \beta)$. However, if the system experiences an imaginary rotation, then the boundary conditions change:
\beqn
\phi({\bs x},\tau) & = & \phantom{-} \phi\bigl({\hat R}({\beta{\bs \Omega}_I}) {\bs x},\tau + \beta\bigr)\,,
\label{eq_phi_rotation}\\
\psi({\bs x},\tau) & = & - \psi\bigl({\hat R}({\beta{\bs \Omega}_I}) {\bs x},\tau + \beta\bigr)\,,
\label{eq_psi_rotation}
\eeqn
where ${\hat R}({\bs\theta})$ denotes a $3 \times 3$ matrix of the spatial rotation, ${\bs x} \to {\bs x}' = {\hat R}({\beta{\bs \Omega}_I}) {\bs x}$, by the angle $\theta \equiv |{\bs \theta}|$ about the axis ${\bf e}_\theta = {\bs \theta}/|\theta|$. For definiteness, we consider the rotation around the $z$ axis, and therefore the boundary condition for the bosonic field~\eq{eq_phi_rotation} can be written in the cylindrical coordinates in the following form:
\beqn
\phi(\rho,\varphi,z,\tau) & = & \phantom{-} \phi(\rho,\varphi - \beta \Omega_I, z,\tau + \beta)\,,
\label{eq_phi_rotation_cyl}\\
\psi(\rho,\varphi,z,\tau) & = & - \psi(\rho,\varphi - \beta \Omega_I, z,\tau + \beta)\,,
\label{eq_psi_rotation_cyl}
\eeqn 
These boundary conditions are visualized in Fig.~\ref{fig_cylinders}. 

We call Eqs.~\eq{eq_phi_rotation}-\eq{eq_psi_rotation_cyl} the ``rotwisted'' boundary conditions (from the combination of the words ``rotation'' and ``twisted''). They share similarities with the ``shifted'' boundary conditions~\cite{Giusti:2010bb} where an imaginary-time translation over the full period is supplemented by a spatial translation. The translationally shifted boundary conditions are related to the generating function of the momentum distribution of fields. They allow, for example, to compute thermodynamic potentials~\cite{Giusti:2011kt} and renormalize the energy-momentum tensor nonperturbatively~\cite{Giusti:2015daa}.

Evidently, the effect of imaginary rotation exhibits the $2 \pi T \equiv 2 \pi / \beta$ periodicity for an expectation value $\cO$ in a bosonic theory,
\beqn
\cO(\Omega_I) = \cO(\Omega_I + 2 \pi n/\beta)\,, \quad \mbox{for} \quad n \in {\mathbb Z}\,,
\label{eq_periodicity}
\eeqn
and the $4 \pi / \beta$ periodicity for the fermionic fields~\cite{Chernodub:2020qah}. Below we concentrate on bosonic theories only. 

The angle of rotation between $\tau = 0$ and $\tau = \beta$ time slices is
\beqn
\theta = [\Omega_I \beta]_{2\pi}\,,
\label{eq_theta}
\eeqn
where
\beqn
[x]_{2\pi} = x + 2 \pi k \in [-\pi,\pi), \qquad k \in {\mathbb Z}\,.
\label{eq_x_2pi}
\eeqn
The angle $\theta$ in Eq.~\eq{eq_theta} is the $2\pi$-periodic function~\eq{eq_periodicity}. While this periodicity is not evident from the light element~\eq{eq_d_s_E}, one notices that the metric in the rotating frame~\eq{eq_d_s_E} can be obtained from the laboratory (static) metric by the identification of the laboratory and corotating variables, respectively: $\varphi_{\mathrm{lab}}(\tau) = \varphi - \Omega_I \tau$. These two types of imaginary rotations, imposed by the curved metric~\eq{eq_d_s_E} and by the boundary conditions~\eq{eq_phi_rotation_cyl} can only be equivalent at low angular frequencies, $\beta |\Omega_I| \ll 2\pi$, where the periodicity in $\Omega_I$ is negligible. Thus the discussed implementations of the imaginary rotation differ from each other. 

The periodicity~\eq{eq_periodicity} also highlights the fact that, contrary to the real rotation $\Omega$, the causality does not restrict the value of the imaginary frequency $\Omega_I$ which can be an arbitrarily large quantity. Indeed, in the imaginary spacetime, there is no light cone (and, consequently, no light cylinder), and the $g_{\tau\tau}$ component never flips its sign in the Euclidean metric~\eq{eq_d_s_E}.

Curiously, the imaginary rotation leads to unusual spin-statistic relations for bosonic and fermionic theories and implies the equivalence of the thermal states for bosonic, fermionic, and exotic ghost fields at certain imaginary frequencies~\cite{Chernodub:2020qah,in_preparation}.

\subsection{Euclidean Tolman-Ehrenfest relation}

In a non-rotating system, $\Omega_I = 0$, the order parameter of the deconfinement phase is the expectation value $\avr{P}$ of the Polyakov loop,
\beqn
{\mathcal P}({\bs x}) = {\mathrm {Tr}}\, {\mathcal P} \exp\left\{ i \oint_{\cC} {\hat A}^a_\tau({\bs x},\tau) d \tau \right\}\,,
\label{eq_P}
\eeqn
where ${\hat A}_\mu \equiv t^a A^a_\mu$ is the $SU(N_c)$ gauge field and $t^a$ with $a = 1, \dots, N_c^2 -1$ are the generators of the $SU(N_c)$ gauge group. The operator $\mathcal P$ stands for the path ordering and the integration takes place along the path $\cC$ directed along the imaginary time direction~$\tau$. The Polyakov loop ${\mathcal P}({\bs x})$ inserts an infinitely heavy static quark $Q$ at the point $\bs x$. Its expectation value is related to the free energy of a single quark $F_Q$ as follows: $F_Q = - T \ln \avr{{\mathcal P}}$.

In an infinite spatial volume, the expectation value of the Polyakov loop $\avr{{\mathcal P}}$ vanishes in the confining phase signaling that the global center ${\mathbb Z}_{N_c}$ symmetry is unbroken and the free energy of an isolated quark is infinite. The Polyakov loop gets a nonzero expectation value in the deconfinement phase where the breaking of the center symmetry occurs.

The operator~\eq{eq_P} is a gauge-invariant quantity because the integration path is closed due to the periodic boundary conditions. Such a path, for example, is given by the segments $O' O$, $PP'$, or $A' A$ in Fig.~\ref{fig_cylinders}(a). The integration path $\cC$ winds once about the (imaginary) time direction. The path is collinear with the time axis which identifies the Euclidean ``Killing vector''~\cite{Rovelli:2010mv} with the imaginary time direction. 

In the (imaginarily) rotating system, a singe-winding Polyakov loop should be defined only in the frame which co-rotates with the matter. For example, in Fig.~\ref{fig_cylinders}(b), the closed loop is given by the curve $P' P$, which represents the world path of the rotating matter in Euclidean spacetime. The corresponding operator, $P_{\cC_{P' P}}$, is a gauge-invariant quantity. On the contrary, at any spatial point ${\bs x}$ different from the origin, ${\bs x} \neq {\bs 0}$, the loop $\cC$ cannot be parallel to the imaginary time vector of the laboratory frame. A relevant example is given by the $A' B$ segment in Fig.~\ref{fig_cylinders}(b) which is not a closed curve due to the rotwisted boundary condition~\eq{eq_phi_rotation} or \eq{eq_phi_rotation_cyl}. Consequently, the corresponding operator, $P_{\cC_{A' B}}$, is a not gauge-invariant quantity and, therefore, cannot serve as an order parameter.

What is the local equilibrium temperature of the rotating matter? In the standard, non-rotating case, the temperature is identified with the inverse length, $T \equiv 1/\beta = 1/L_\tau$, of the compactified time $S^1_\tau$. The quantity $L_\tau$ corresponds to the length of the imaginary path of a static quark. For a quark that resides in the (imaginary) rotating matter, this path is given by the shortest trajectory of a heavy particle rotating together with matter (plasma). Such a trajectory (worldline) fulfills the rotwisted boundary conditions~\eq{eq_phi_rotation_cyl}. A relevant example is given by the segments $A' A$ or $P' P$ in Fig.~\ref{fig_cylinders}(b). 

The length of the worldline of a rotating heavy quark in the Euclidean space is $L_\tau(\rho) = \sqrt{\beta^2 + \theta^2 \rho^2}$, where the angle $\theta$ is given in Eq.~\eq{eq_theta}. Therefore, the heavy quark finds itself immersed in a rotating heat bath with the temperature $T^E_{\mathrm{TE}} = 1/L_\tau(\rho)$, or
\beqn
T^E_{\mathrm{TE}}(\rho, \Omega_I) = \frac{T_0}{\sqrt{1 + \rho^2 T^2_0 [\Omega_I/T_0]_{2\pi}^2}}\,,
\label{eq_T_rho_E}
\eeqn
where the superscript ``E'' indicates that Eq.~\eq{eq_T_rho_E} generalizes the TE relation to the imaginary-time formalism in the Euclidean space. The quantity $T_0 \equiv T^E_{\mathrm{TE}}(0, \Omega_I)$ has the sense of temperature at the axis of imaginary rotation, at $\rho = 0$. Notice that the Euclidean TE temperature~\eq{eq_T_rho_E} possesses the periodicity~\eq{eq_periodicity} with respect to the shifts $T^E_{\mathrm{TE}}(\rho, \Omega_I) = T^E_{\mathrm{TE}}(\rho, \Omega_I + 2 \pi T)$.

For imaginary rotations in the elementary domain of angular frequencies, the Euclidean TE relation~\eq{eq_T_rho} reduces to 
\beqn
T^E_{\mathrm{TE}}(\rho, \Omega_I) = \frac{T_0}{\sqrt{1 +  \rho^2 \Omega_I^2}}, \quad \mbox{for} \quad |\Omega_I| < \pi T_0, \quad
\label{eq_T_rho_E_reduced}
\eeqn
which, after the analytical continuation~\eq{eq_Omega_I_2} to Minkowski space, leads us the familiar TE relation~\eq{eq_T_rho}. Given this relation, the quantity~\eq{eq_T_rho_E} can be interpreted as the temperature of the heat bath as observed by the heavy quarks (particles, in general) that co-rotate together with the plasma. 

Notice that Eq.~\eq{eq_T_rho_E} has a kinetic nature with no dynamical arguments involved in the derivation. This relation can be considered as a consequence of a redshift effect in the Euclidean geometry subjected to the imaginary rotation. In addition, there is no restriction on the imaginary angular velocity $\Omega_I$, which could appear as a result of causality. There is no light cone and no causality in Euclidean spacetime.  On the contrary, in Minkowski spacetime, the TE temperature becomes imaginary~\eq{eq_T_rho} at $|\Omega| > R^{-1}$ highlighting the physical need to impose the causality constraints. 

\subsection{Phase structure at imaginary rotation}

The Euclidean TE relation~\eq{eq_T_rho_E} suggests a particular structure for the phase structure of rotating Euclidean Yang-Mills theory (and also of rotating QCD since the above relation has a kinetic nature). The sketch of the phase diagram for a fixed $\Omega_I$ in a cylinder of a fixed radius $R$ is shown in the lower panel of Fig.~\ref{fig_phaselines}. 

Since the local temperature~\eq{eq_T_rho_E} is a decreasing function of the radius $\rho$, then the first critical point is given by the transition temperature in the infinite volume, $T_{c,\infty}$:
\beqn
T^E_{c1} = T_{c,\infty}\,.
\label{eq_Tc_1_E}
\eeqn
Indeed, if the temperature of the system in the center of rotation $T_0$ is lower than $T_{c,\infty}$, then the whole volume resides in the confinement phase. The first critical temperature for imaginary rotation~\eq{eq_Tc_1_E} corresponds to the second critical temperature for the real rotation~\eq{eq_Tc_1}.

If the temperature in the center exceeds the first critical temperature~\eq{eq_Tc_1_E} for imaginary rotation, then the center experiences a deconfining transition. However, as we move further from the axis of rotation, the temperature drops down, and the system enters the confining phase again. Therefore, right above the first critical temperature~\eq{eq_Tc_1_E}, we have an inhomogeneous phase in Euclidean space similar to the mixed phase of the rotating gluon gas in Minkowski spacetime (shown in the lower panel of Fig.~\ref{fig_phaselines}) with, however, confining and deconfining phases swapped. 

For a cylinder of a finite radius $R$, the whole space becomes deconfined if the temperature exceeds the second critical point:
\beqn
T^E_{c2} = T_{c,\infty} \sqrt{1 + \Omega^2_I R^2}\,, \quad \mbox{for} \quad -\pi \leqslant \Omega_I \beta < \pi.\quad
\label{eq_Tc_2_E}
\eeqn
For the sake of simplicity,  we restricted the imaginary angular velocity $\Omega_I$ to the elementary segment. The generalization to other $\Omega_I$, because of periodicity~\eq{eq_periodicity}, is straightforward. 

In summarizing, the gluon gas in the cylindrical geometry, rotating with imaginary angular frequency in Euclidean space, possesses the phase diagram shown in the lower panel of Fig.~\ref{fig_phaselines}. The critical temperatures are given by Eqs.~\eq{eq_Tc_1_E} and \eq{eq_Tc_2_E}. For the same geometry, the real rotation in Minkowski spacetime provides us with the phase diagram shown in the upper panel of the same figure, with critical temperatures given by Eqs.~\eq{eq_Tc_1} and \eq{eq_Tc_2}. These phase diagrams are related to each other by the analytical continuation~\eq{eq_Omega_I_2} which maps the corresponding TE laws, \eq{eq_T_rho} and \eq{eq_T_rho_E_reduced}, respectively. This one-to-one matching allows us to use the Euclidean simulations to probe the structure of the rotating gluon gas in Minkowski spacetime. For instance, the observation of the inhomogeneous confinement-deconfinement phase in the imaginary-rotating Euclidean gluon plasma at $T^E_{c1} < T < T^E_{c2}$ implies the existence of the inhomogeneous mixed phase of gluons rotating in Minkowski spacetime at the range of temperatures $T_{c1} < T < T_{c2}$. 

Coming closer to the simulations of lattice Yang-Mills theory, we remind that the rotation in Euclidean space has no causality requirements. Therefore there is no physical reason to bound the Euclidean lattice system in a cylindrical geometry (which would make our simulations more difficult). Consequently, we consider the Euclidean system in a large spatial volume sending, effectively, the radius of the cylinder $R$ to infinity. At the practical, the second critical temperature~\eq{eq_Tc_2_E} becomes large (infinite), and the phase diagram of rotating Yang-Mills gas acquires the two-phase structure with the single critical temperature~\eq{eq_Tc_1_E} which separates the low-temperature confinement phase with the high-temperature mixed confinement-deconfinement phase. The expected phase diagram in the $(T,\Omega_I)$ plane is shown in Fig.~\ref{fig_phase_diagram}(b).

\begin{figure}[ht]
\centering
  \includegraphics[width=0.95\linewidth]{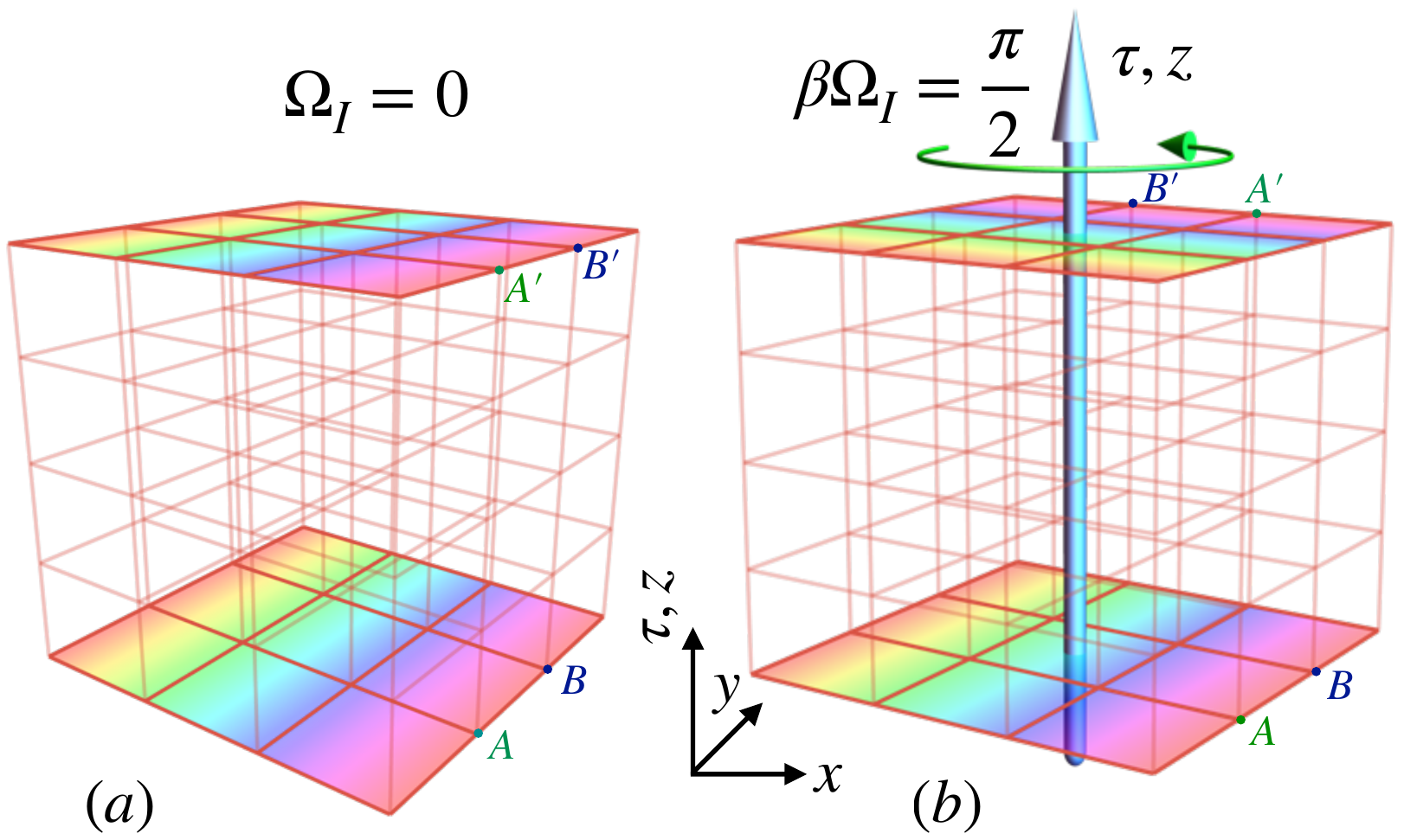}
  \caption{The hypercubic Euclidean lattice: (a) the periodic boundary conditions at vanishing imaginary angular frequency $\Omega_I = 0$: the points $A$ and $A'$ as well as, respectively, $B$ and $B'$, are pairwise identified. (b) The rotwisted boundary conditions imposed on the quarter-rotated lattice with $\pi/2$ angle~\eq{eq_quarter} with the identification of the temporal boundaries~\eq{eq_Pi2_matching}. As in Fig.~\ref{fig_cylinders}, the axis $z$ and $\tau$ are shown symbolically: as the imaginary time advances for the full period $\tau \to \tau + \beta$ along the compactified direction $S^1_\tau$, the $xy$ plane either (a) does not rotate for the periodic conditions or (b) rotates for the rotwisted conditions at the angle $\pi/2$ along the compactified direction $S^1_\tau$. The identification of the link and plaquettes is evident from the global geometrical orientation of the $xy$ planes shown by the coloring.}
  \label{fig_lattices}
\end{figure}

\section{Imaginary rotation on the lattice}
\label{sec_lattice}

\subsection{Quarter-imaginary-rotation on the lattice}

Can we implement the imaginary rotation~\eq{eq_phi_rotation_cyl} in the standard hypercubic geometry, which is used for simulations of lattice gauge theories? Yes, this is possible, but the rotation angle $\beta \Omega_I$ must be consistent with the lattice symmetries dictated by the underlying $C_4$ rotation group. In other words, after a full imaginary time period, $\tau\to \tau + \beta$, the system can rotate only at quarter-quantized angles $\beta\Omega_I = (\pi/2) k$ with $k \in {\mathbb Z}$. The periodicity of the imaginary rotation implies that the only plausible choice of the angular frequency is as follows:
\beqn
\Omega_I = \frac{\pi}{2} T\,,
\label{eq_quarter}
\eeqn
which is visualized in Fig.~\ref{fig_lattices}. According to Eq.~\eq{eq_phi_rotation_cyl}, the imaginary counterclockwise rotation~\eq{eq_quarter} corresponds to the matching at the boundaries:
\beqn
(x,y,z,\tau) \to  (-y,x,z,\tau + \beta)\,.
\label{eq_Pi2_matching}
\eeqn
Another possible option of the imaginary frequency on the hypercubic lattice, $\Omega_I = \pi T$, corresponds to a half-rotation, 
\beqn
(x,y,z,\tau) \to  (-x,-y,z,\tau + \beta)\,,
\eeqn
which is neither clockwise nor counterclockwise. The same proposal to consider the imaginary frequency~\eq{eq_quarter} has also been put forward in Ref.~\cite{Chen:2022smf} (see also the acknowledgments). 

An illustration of the $\pi/2$ rotwisted boundary condition~\eq{eq_quarter} in comparison with the standard periodic boundary condition is shown in Fig.~\ref{fig_lattices}. 

The imaginary rotation with the angular frequency~\eq{eq_quarter} corresponds to a very fast rotation which might question the validity of (even formal) analytical continuation~\eq{eq_Omega_I_2} from the Euclidean space to Minkowski spacetime. Moreover, the choice~\eq{eq_quarter} should generate lattice artifacts stipulated by the $C_4$ lattice group. However, our article aims to perform an exploratory study of the qualitative effects of rotation properties of thermal Yang-Mills plasma, including a possible indication of the two-phase structure. Therefore, we concentrate below on the fast imaginary rotation with the single value of the imaginary frequency~\eq{eq_quarter}.

\begin{figure}[!thb]
\centering
  \includegraphics[width=0.985\linewidth]{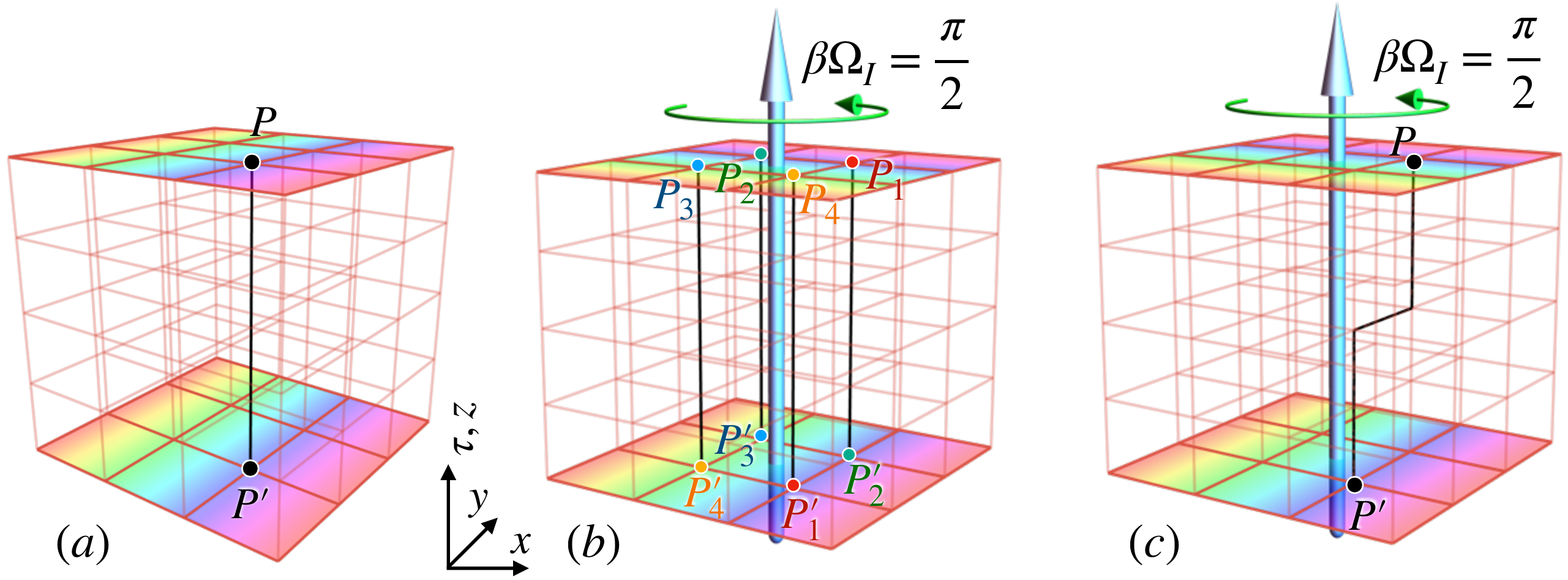}
  \caption{The Polyakov loop order parameters: (a) elementary, single-fold winding loop ${\mathcal P}$ for a static ($\Omega_I = 0$) non-rotating plasma with $P$ and $P'$ denoting the same point identified via the periodic boundary condition. A rapidly rotating plasma with the $\pi/2$ imaginary frequency~\eq{eq_quarter}: (b) a fourfold all-straight closed Polyakov loop ${\mathcal P}_4$ in the laboratory frame with the points $P_a$ and $P_a'$ ($a = 1, \dots, 4$) pair-wise identified; (c) single-fold winding loop ${\mathcal P}$ in the co-rotating frame, with a spacelike jumper that allows to match the points $P$ and~$P'$. The latter lattice contour is a lattice analog of the $P' P$ closed curve in Fig.~\ref{fig_cylinders}(b).}
  \label{fig_loops}
\end{figure}

\subsection{Polyakov order parameter in rotating matter}

The phase structure of the theory can be revealed with the help of the relevant order parameters. In the case of the thermal deconfining transition in Yang-Mills theory, the order parameter is the Polyakov loop~\eq{eq_P} which is closed via the temporal boundary conditions. 

In the standard, non-rotating case, the Polyakov loop is a straight line parallel to the imaginary time direction, as shown in Fig.~\ref{fig_loops}(a). For the $\pi/2$-rotwisted boundary condition~\eq{eq_quarter} with the boundary matching~\eq{eq_Pi2_matching}, there are two possible choices for the Polyakov loop:
\begin{itemize}
    \item One can consider a static four-fold loop ${\mathcal P}_4$ of the length $4 L_\tau$ which pierces the lattice four times thus making a full, $2\pi = 4 \times \frac{\pi}{2}$ angle, Fig.~\ref{fig_loops}(b):
    \beqn
    {\mathcal P}_4 = {\mathrm{Tr}}\, U_{P_1'P_1} U_{P_2'P_2} U_{P_3'P_3} U_{{\mathcal P}_4'{\mathcal P}_4}\,,
    \label{eq_P4}
    \eeqn
    where $U_{P_a' P_a}$ indicated the segment of the ordered product of the elementary link matrices $U_l$ along the straight line $P_a' P_a$. The operator ${\mathcal P}_4$ corresponds to the order parameter defined in the laboratory frame, which does not rotate with the thermal matter. The latter reason leads to suspicion, from the very beginning, that this four-fold operator should be irrelevant for the thermal transition. 

    \item One can also define a single-winding loop ${\mathcal P}$ shown in Fig.~\ref{fig_loops}(c): 
    \beqn
    P = {\mathrm{Tr}}\, U_{P' P}\,.
    \label{eq_P_jumper}
    \eeqn
    This loop has a ``jumper'' segment in the spatial space, which makes it closed given the identification~\eq{eq_Pi2_matching} of the points $P$ and $P'$ via the rotwisted boundary conditions. The discrete loop $P' P$ in Fig.~\ref{fig_loops}(c) is the lattice version of the loop $P' P$ in the continuum theory, Fig.~\ref{fig_cylinders}(b).

\end{itemize}

Both laboratory-frame (four-fold) Polyakov loop ${\mathcal P}_4$ and co-rotating-frame (single-fold) Polyakov loop ${\mathcal P}$ are sensitive to the center ${\mathbb Z}_3$ symmetry, respectively,
\beqn
{\mathcal P} \to e^{\frac{2 \pi i}{3} n} {\mathcal P}\,, \qquad 
{\mathcal P}_4 \to e^{4 \frac{2 \pi i}{3} n} {\mathcal P}_4\,, \qquad n = 0,1,2,\quad 
\label{eq_Z3_transformation}
\eeqn 
as both transform nontrivially for the non-unit elements (with $n=1,2$) of the center group.

Below, we present the numerical results for all three physical cases shown in Fig.~\ref{fig_loops}. 

\subsection{Numerical results}

\subsubsection{Setup}

We simulate the $SU(3)$ gauge model with the standard Wilson action
\beqn
S = \beta \sum_{P} \left( 1 - \frac{1}{3} {\mathrm{Re}}\, {\mathrm{Tr}}\, U_P \right)\,,
\eeqn
on the lattices $N_s^3 \times N_\tau$ with the temporal extension $N_\tau = 8$ and the spatial sizes $N_s = 32$ and $48$. We use the standard heatbath formalism~\cite{Gattringer:2010zz} to generate $10^5$ configurations at each numerical measurement. The physical value of the lattice spacing, as a function of the lattice coupling, $a = a(\beta)$, is taken from Ref.~\cite{Edwards:1997xf}. The $SU(3)$ lattice Yang-Mills theory experiences a weak first-order deconfining phase transition at the critical lattice coupling $\beta_c = 6.0609(9)$ at the $32^3 \times 8$ lattice~\cite{Boyd:1996bx}.

\subsubsection{Polyakov order parameter in the co-rotating frame}

We start our discussion from the most exciting case of the Polyakov loop ${\mathcal P}$ defined in the co-rotating frame as shown in Fig.~\ref{fig_loops}(c). To simulate the imaginary rotation, we use (i) the $\pi/2$-rotwisted boundary conditions~\eq{eq_Pi2_matching} -- visualized in Fig.~\ref{fig_lattices}(b) -- for the imaginary time direction $\tau$; (ii) the periodic boundary conditions along the axis of rotation in the direction $z$; and (iii) open boundaries for $x$ and $y$ directions in the transverse, to the rotation axis, plane. The size of the lattice $N_s$ in the $a$ and $y$ directions is determined by the number of the lattice sites. Therefore, for even $N_s$, the geometrical center of the lattice is located in the center of a plaquette, as shown in Fig.~\ref{fig_loops}.

\begin{figure}[ht]
\centering
  \includegraphics[width=0.95\linewidth]{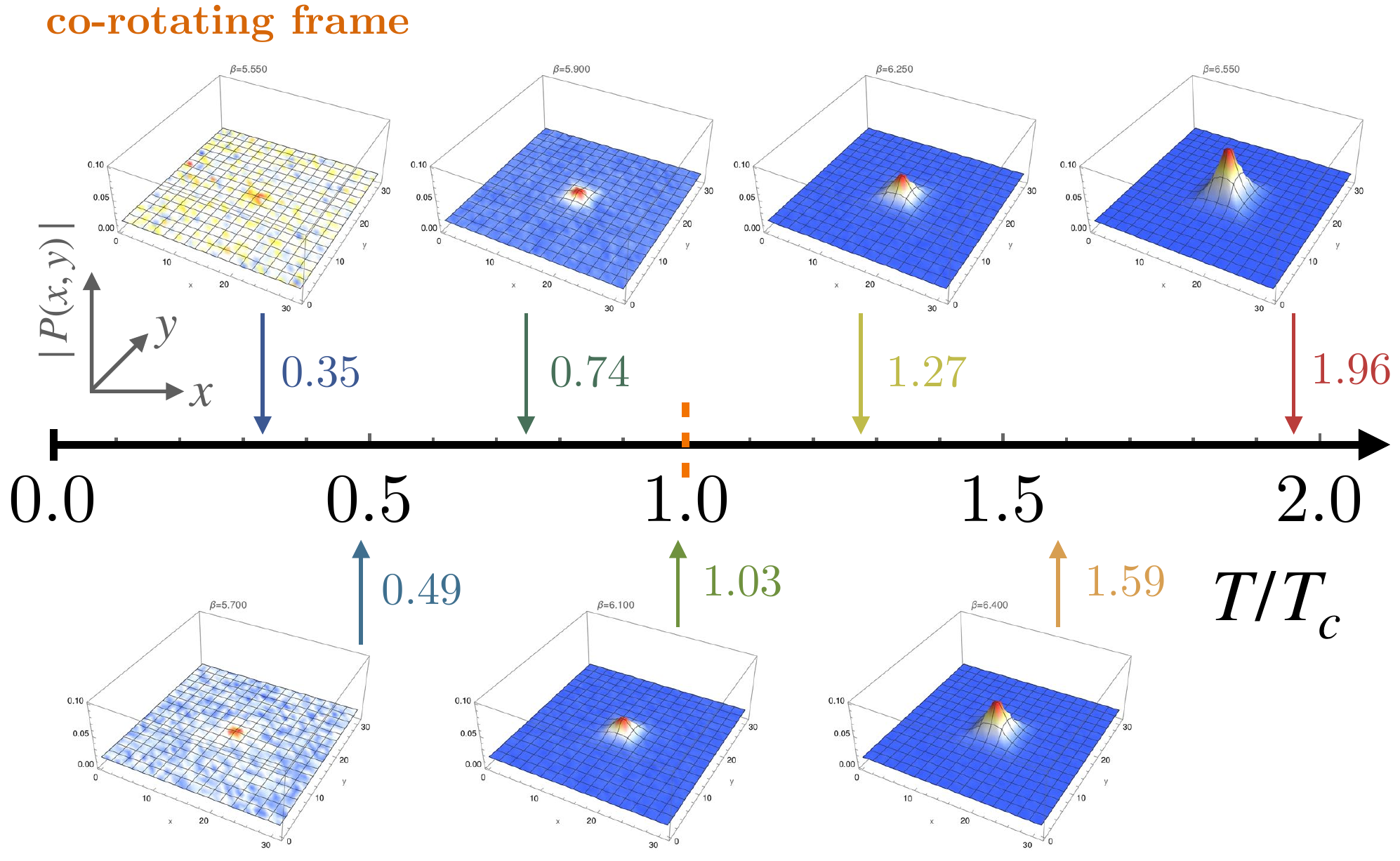}
  \caption{The expectation value of the local Polyakov loop ${\mathcal P}$ calculated numerically in finite-temperature $SU(3)$ gauge theory under the $\Omega_I = \pi T/2$ imaginary rotation. The loop is defined in the co-rotating frame as illustrated in Fig.~\ref{fig_loops}(c). The expectation values are shown in the $(x,y)$ plane normal to the axis of rotation. In temporal direction, the rotwisted boundary condition~\eq{eq_phi_rotation_cyl} corresponding to the $\pi/2$ imaginary rotation~\eq{eq_quarter} is implied. The temperature $T$ and the position of the phase transition $T = T_c$ (the orange arrow) correspond to the non-rotating thermal gluon plasma, $T = 1/L_\tau$.}
  \label{fig_Polyakov_Nt_3d}
\end{figure}

In Fig.~\ref{fig_Polyakov_Nt_3d}, we show the numerical results for the spatial structure of the expectation value of the co-rotating Polyakov loop ${\mathcal P}$ in the $(x,y)$ plane perpendicular to the axis of rotation $z$. The local value of the loop, defined at a spatial point $(x,y,z)$, is averaged along the $z$ axis. 

Our theoretical expectations, based on the Euclidean TE law~\eq{eq_T_rho_E_reduced}, suggest that the rotation creates an inhomogeneous phase (with the phase diagrams shown in Figs.~\ref{fig_phaselines} and \ref{fig_phase_diagram}). Our numerical results, presented in Fig.~\ref{fig_Polyakov_Nt_3d}, indeed show the emergence of the expected inhomogeneity: the hot plasma domain appears close to the center of rotation. The gluonic plasma is surrounded by cold, confining region [{\it cf.} Figs.~\ref{fig_phase_diagram}(a) and \ref{fig_phase_diagram}(b)]. The observation of the inhomogeneous plasma structure in Euclidean spacetime implies the validity of the kinetic TE picture and points out the existence of the rotation-induced inhomogeneous phase in the Minkowski spacetime. Notice that after the Wick transformation, the positions of the confining and deconfining phases are reversed in the inhomogeneous phase, Fig.~\ref{fig_phaselines}.

There are a few technical remarks about Fig.~\ref{fig_Polyakov_Nt_3d}. First, the plasma has a square shape as a reminder of the discrete $C_4$ group of lattice rotations. The effect is enhanced by strong (imaginary) rotation with the large imaginary frequency~\eq{eq_quarter} produced by the $\pi/2$ rotwisted boundary conditions along the compactified (temperature) direction. Second, the signatures of the inhomogeneous plasma appear already in the confining phase. This observation is a consequence of a finite-volume effect since we take the mean of the {\it local} Polyakov loop over the small volume $1 \times 1 \times N_s$. We have two sources of finite-volume effects related to (i) finite total lattice volume, $N_s^3$, and (ii) the finite extension $N_s$ of the $z$ axis available for taking the mean of the local Polyakov loop.

\begin{figure}[ht]
\centering
  \includegraphics[width=0.95\linewidth]{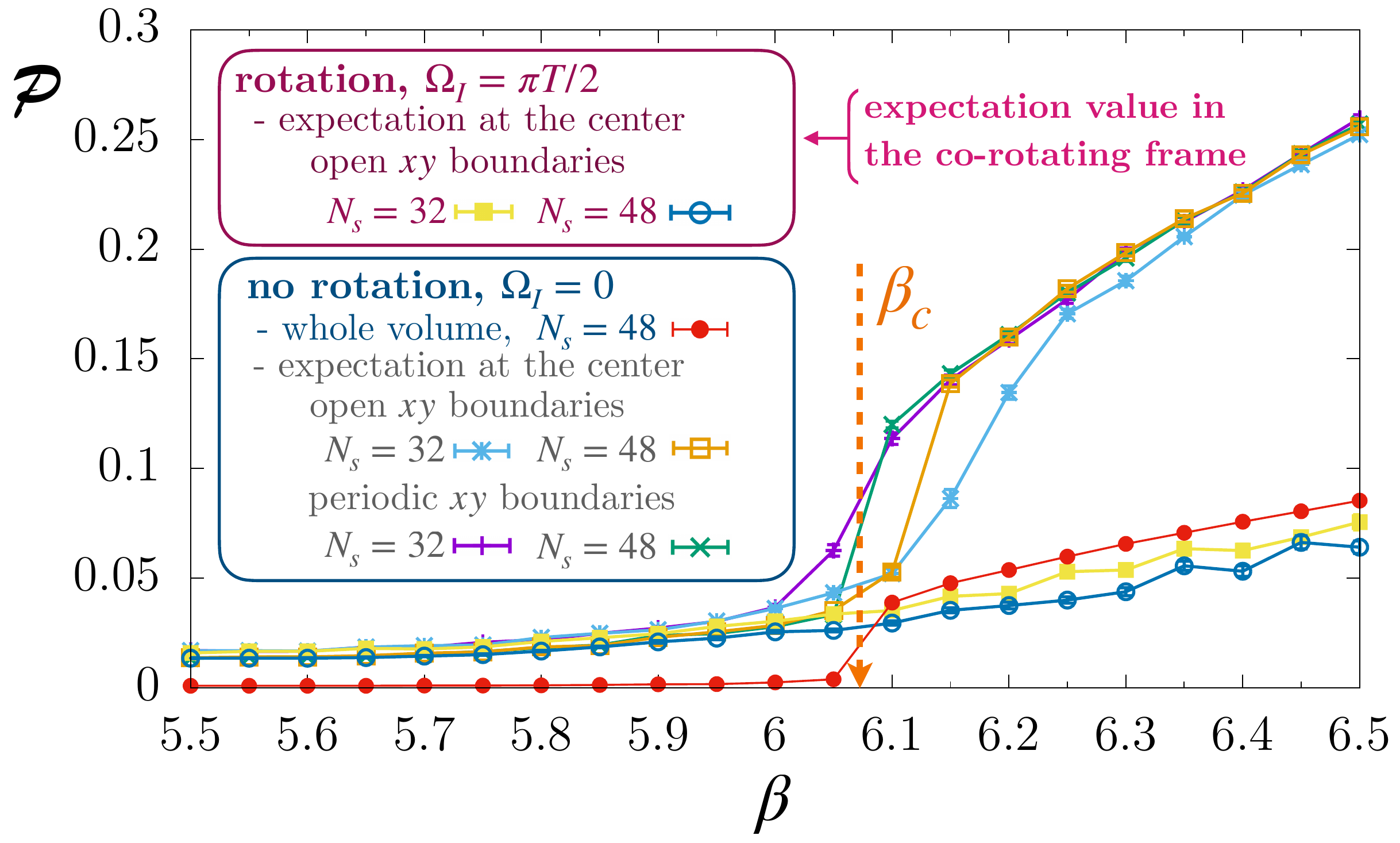}
  \caption{Local expectation value of the co-rotating Polyakov loop ${\mathcal P}$ in the bulk (for the non-rotating lattice at $\Omega_I=0$ only) and at the center of rotation for open and periodic (for the $xy$ plane) boundary conditions for the lattices with $N_s = 32$ and $N_s = 48$ spatial sizes. The dashed orange line shows the critical coupling $\beta_c \equiv \beta_c(\Omega_I = 0)$ of the thermodynamic bulk phase transition for a non-rotating lattice.}
  \label{fig_Polyakov_Nt}
\end{figure}

In Fig.~\ref{fig_Polyakov_Nt}, we show the expectation value of the co-rotating Polyakov loop at the center of rotation\footnote{Notice that since the actual axis of rotation is located at the center of the center plaquette, the loop is placed at the distance $\sqrt{2} a$ from the rotation axis. Since this distance is very short, we identify the Polyakov loop piercing the sites of the center $xy$ plaquette with the ``on-axis'' Polyakov loop.}. The data for the lattices with the spatial sizes $N_s = 32$ and $N_s = 48$ are almost superimposed on each other, so we conclude that the effects imposed by the finite spatial lattice volume are small. 

In the deconfining phase, the expectation value of the co-rotating {\it local} Polyakov loop at the rotation axis (with $\Omega_I = \pi T/2$) approaches the expectation value of the volume-averaged (bulk) Polyakov loop calculated in the non-rotating lattice (with $\Omega_I = 0$). Thus, above the deconfining temperature $T > T_c^{\infty}$, the gluons at the center of the imaginary rotation reside in the deconfinement phase (it is not a trivial fact since, at the same time, at $T > T_c^{\infty}$, the gluons far from the rotation axis appear to be in the confining phase as it is seen from Fig.~\ref{fig_Polyakov_Nt_3d}).

As temperature diminishes, the bulk Polyakov loop at $\Omega_I = 0$ rapidly vanishes, while the co-rotating loop at $\Omega_I = \pi T/2$ approaches very slowly a non-vanishing expectation value. The smoothness of the $\Omega_I = \pi/2$ transition could be a physical effect since in the non-rotating case, with $\Omega_I = 0$, the on-axis Polyakov loop behaves much sharper. The data for the latter quantity, shown in the same Fig.~\ref{fig_Polyakov_Nt}, shows qualitative insignificance of the type of the boundary conditions (open vs periodic) and exhibits insensitivity to the spatial lattice size ($N_s = 32$ vs $N_s = 48$).

The non-vanishing value of co-rotating is a finite-volume effect of the second type related to the locality of the on-axis Polyakov loop. This fact is seen from the coincidence of the values of the co-rotating loop at $\Omega_I = \pi T/2$ with the results for the local Polyakov loop for the non-rotating lattice in Fig.~\ref{fig_Polyakov_Nt}.

The spatial structure of the rotating Yang-Mills theory revealed in this paper with the help of rotwisted boundary conditions at $\Omega_I \sim T_c$, differs from the results obtained with the help of the curved rotational metric imposed on the lattice at low angular frequencies~$\Omega_I \ll T_c$~\cite{Braguta:2020biu,Braguta:2021jgn}. In the latter case, no pronounced inhomogeneity at the center is seen: the local order parameter, the Polyakov loop, depends on the spatial boundary conditions near the system's edges but not in bulk. The origin of this discrepancy can be related to the qualitative difference in the implementation of imaginary rotation (rotwisted boundary condition vs. curved lattice spacetime) as well as be ascribed to the quantitative factor: our imaginary frequency is the order of magnitude higher than the one imposed in Refs.~\cite{Braguta:2020biu,Braguta:2021jgn}.

We also do not see a clear deconfinement transition for the rotating gas at an elevated temperature $T_c(\Omega_I = \pi T/2) > T_c(\Omega_I = 0)$ as it was suggested recently in Ref.~\cite{Chen:2022smf}. On the contrary, our data for the rotating gas, reported in Fig.~\ref{fig_Polyakov_Nt}, shows a slow rise of the on-axis Polyakov loop starting in the confinement phase. One could tempt to attribute the smooth behavior of the on-axis Polyakov loop to a finite-volume effect since we average the local Polyakov loop in a small spatial volume  $N_s \times 1^2$ along the $z$ axis. However, as we mentioned above, the same local Polyakov loop in the non-rotating Yang-Mills theory shows a clear signature of the phase transition thus excluding the finite-volume argument. Therefore, our data suggests that the Euclidean Yang-Mills plasma rotating with the imaginary frequency $\Omega_I = \pi T/2$ possesses the single crossover-type deconfinement transition in the spatial region close to the rotating axis at the temperature close to the critical deconfining temperature of the non-rotating gas. 

\subsubsection{Rotwisted boundary condition and physical temperature}

One could also mention another intriguing possibility by suggesting that the rotwisted boundary condition, imposed along the imaginary time direction, modifies the temperature in the whole lattice. Analysis of classical solutions shows that the system subjected to the imaginary rotation with the rational angular momentum $\Omega_I/(2 \pi T) =  p/q$ with natural nonzero numbers $p,q\in {\mathbb N}$, could be exposed to the thermal bath with following the laboratory-frame temperature~\cite{Chernodub:2022wsw}:
\beqn
T_{\mathrm{lab}} = \frac{1}{q L_\tau}, \qquad \Omega_I = \frac{2 \pi}{T} \frac{p}{q} \,,
\label{eq_T_lab}
\eeqn
where $L_\tau \equiv a N_\tau$ is the length of the lattice in the imaginary-time direction. For the $\pi/2$-rotation, the physical temperature should be lower by the factor of four, $T_{\mathrm{lab}} = T_0/4 \equiv 1/(4 L_\tau)$ as compared both to the naively computed temperature $T_0 = 1/L_\tau$ and to the Euclidean TE temperature~\eq{eq_T_rho_E}. This factor is consistent with the four-fold increase in the length of the Polyakov loop ${\mathcal P}_4$, Fig.~\ref{fig_loops}(b), with respect to the length of the compact time dimension, Fig.~\ref{fig_loops}(b). Therefore, our results for the deconfining order parameter at the non-zero $\Omega_I$, shown in Fig.~\ref{fig_Polyakov_Nt}, could also formally refer to a confining phase of the theory from the point of view of the laboratory-defined operator ${\mathcal P}_4$.

In order to check the qualitative validity of our results, we extended our calculations to very large lattice couplings, $\beta = 10$. We consistently observed the presence of the inhomogeneous, two-phase structure in the whole domain of studied couplings. Moreover, the coincidence of the on-axis Polyakov loop in the rotating frame (the blue dataset in Fig.~\ref{fig_Polyakov_Nt}) and the bulk Polyakov loop of the non-rotating lattice (the red dataset in the same figure) extends to all measured points in the deeply deconfining domain (we checked the validity of this coincidence till $\beta = 7$). Thus, the center of the imaginarily rotating plasma in the deconfinement phase ($T>T_c$) stays in the deconfinement phase while the exterior is confining. 

We do not report in the present paper the concrete data for large values of lattice coupling $\beta$ because they bring strong finite-volume effects that should spoil the quantitative validity of the results. Indeed, at large $\beta$, the spatial size of our lattices becomes so small that the system experiences the unphysical finite-volume transition to the deconfinement phase. This question requires an additional careful investigation.

Thus, the imaginary rotation in Euclidean space generates the inhomogeneous structure of the gluon plasma consisting of the hot plasma domain close to the rotation axis surrounded by the confining phase in bulk. In the confining phase, the central plasma domain disappears, and the confining phase occupies the whole volume. This result in Euclidean space supports the Minkowski picture shown in Fig.~\ref{fig_phaselines}.

\subsubsection{Polyakov order parameter in the laboratory frame}

\begin{figure}[ht]
\centering
  \includegraphics[width=0.95\linewidth]{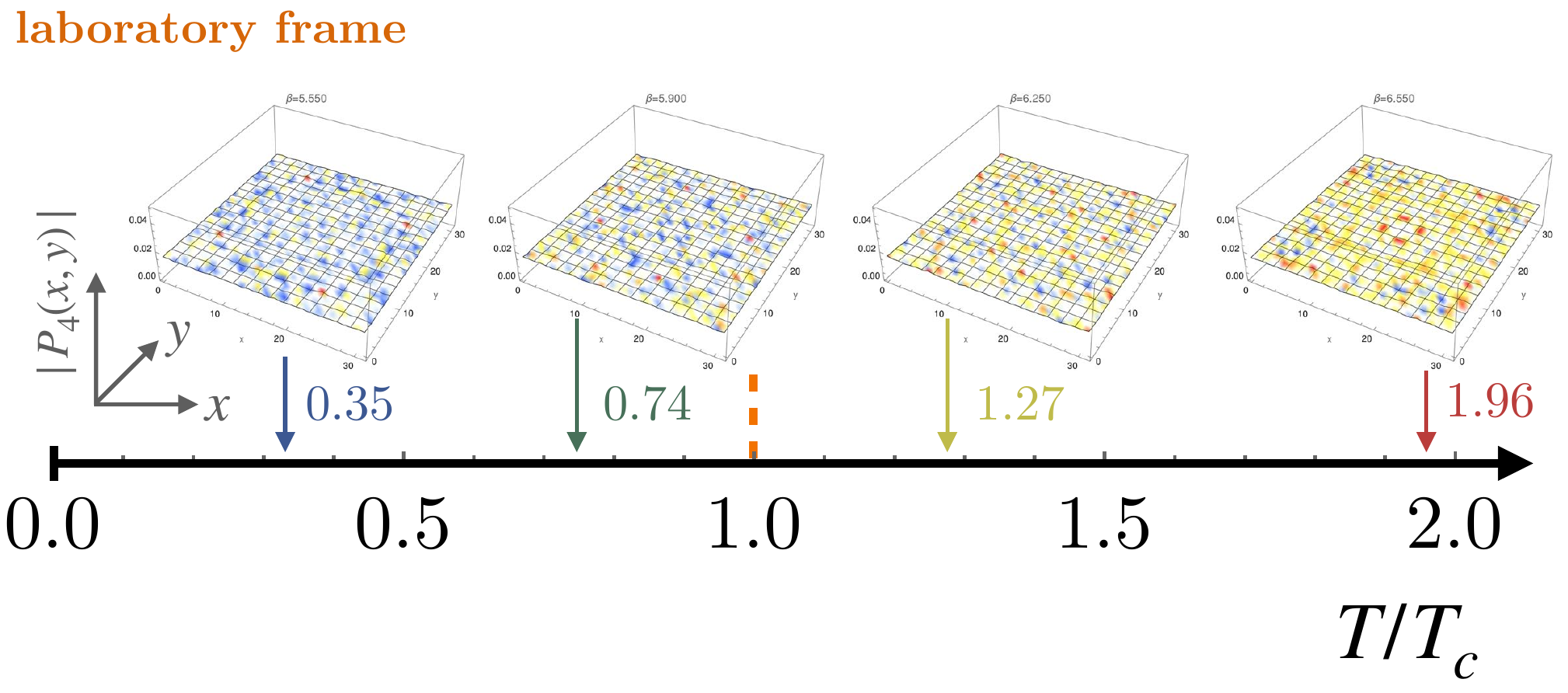}
  \caption{The same as in Fig.~\ref{fig_Polyakov_Nt_3d} but for four-fold Polyakov loop ${\mathcal P}_4$, Fig.~\ref{fig_loops}(b), defined in the laboratory frame at the imaginary angular frequency $\Omega_I = \frac{\pi}{2} T$.}
  \label{fig_Polyakov_4Nt_3d}
\end{figure}

In the laboratory frame of the $\Omega_I = \pi T/2$ rotation, the Polyakov loop ${\mathcal P}_4$ winds four times about the imaginary time axis, as illustrated in Fig.~\ref{fig_loops}(b). Our Monte Carlo results for the spatial distribution of the expectation value of this loop are shown in Fig.~\ref{fig_Polyakov_Nt_3d}. The four-fold loop ${\mathcal P}_4$ appears to be relatively insensitive to temperature. Moreover, contrary to the co-rotating Polyakov loop $\mathcal P$, Fig.~\ref{fig_Polyakov_4Nt_3d}, no spatial structure in its expectation value of ${\mathcal P}_4$ is seen in the laboratory frame. 

\begin{figure}[ht]
\centering
  \includegraphics[width=0.95\linewidth]{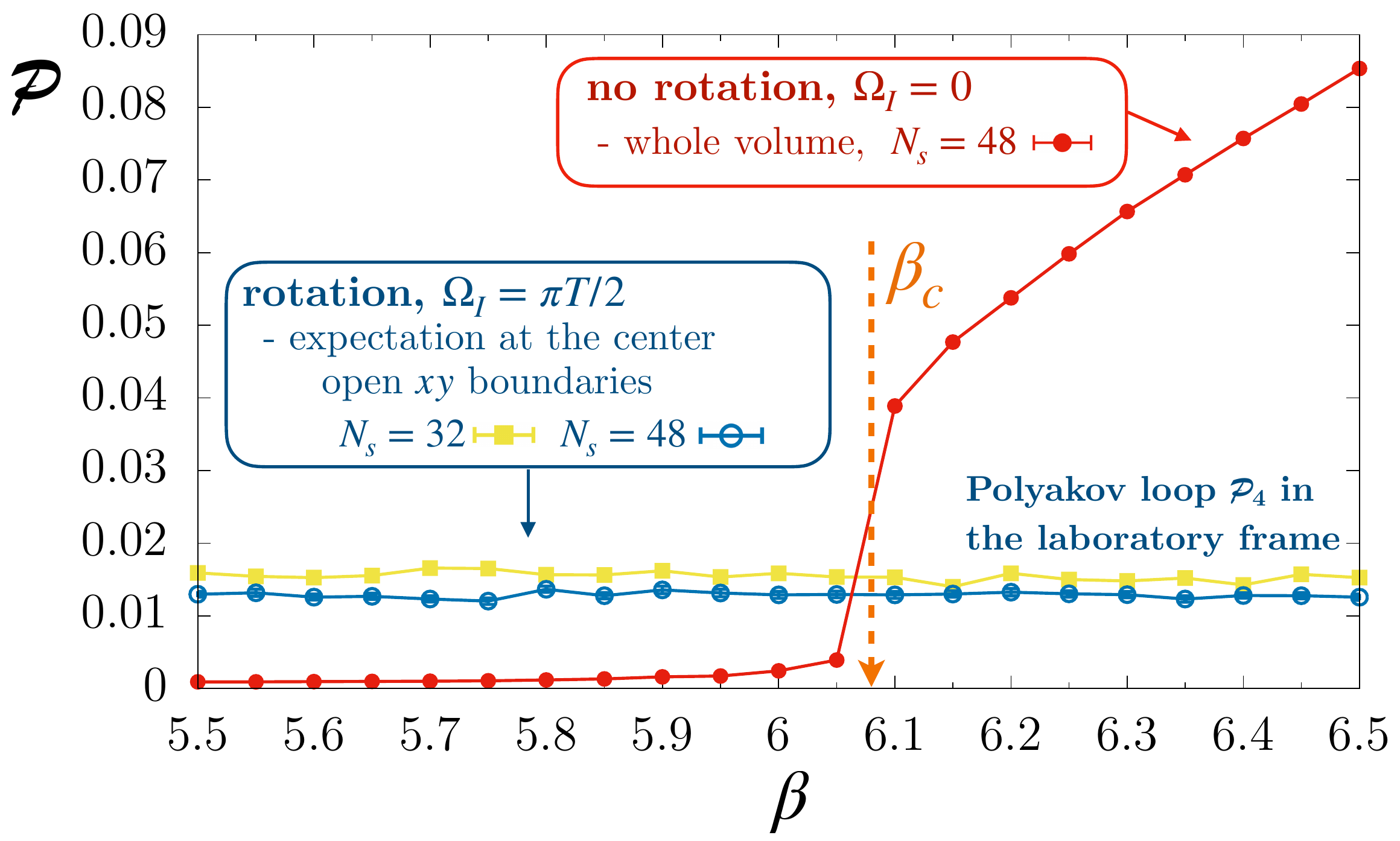}
  \caption{Expectation value of the four-fold Polyakov loop ${\mathcal P}_4$ in the laboratory frame under the imaginary $\Omega_I = \frac{\pi}{2} T $ rotation (the blue and yellow lines) compared with the bulk expectation value of the standard Polyakov loop $P$ in the non-rotating SU(3) gauge theory (the red line). The dashed orange arrow marks the phase transition in the non-rotating case).}
  \label{fig_Polyakov_4Nt}
\end{figure}

Our numerical results for the on-axis expectation value of the four-fold Polyakov loop ${\mathcal P}_4$ are shown in Fig.~\ref{fig_Polyakov_4Nt}. Despite the four-fold loop being sensitive to the center ${\mathbb Z}_3$ symmetry~\eq{eq_Z3_transformation}, no signature of the ${\mathbb Z}_3$ symmetry breaking is observed from the expectation value of this operator in the explored region of lattice couplings $\beta$.\footnote{As for the larger $\beta$ region, we found that the bulk expectation value of the four-fold Polyakov loop ${\mathcal P}_4$ stays at the low constant value till the lattice coupling reaches $\beta \sim 9$. Above this coupling, the loop ${\mathcal P}_4$ starts to rise. This behavior indicates, most probably, the presence of the deconfinement transition caused by the finite physical volume of the system. Should any physical thermodynamics transition exist for the rotwisted lattice at high temperatures, it will be overshadowed by this finite-volume effect for our lattice volumes.} We remind, however, that this operator has a rather questionable theoretical interpretation since it is defined in the laboratory frame of the rotating system. Since the ${\mathcal P}_4$ operator does not introduce a heavy test quark co-rotating with the gluon plasma, its expectation value does not coincide with the free energy of the heavy quark in the rotating plasma.

\section{Conclusion and Discussion}

We studied the properties of hot Euclidean Yang-Mills plasma subjected to rotation with imaginary angular frequency. Such rotation is supported by the ``rotwisted'' (originated from merging the words ``rotation'' and ``twisted'') boundary conditions in the imaginary-time formalism in a finite-temperature field theory. The rotwisted boundary condition is a generalization of the usual periodic (anti-periodic) boundary condition imposed on bosonic (fermionic) fields in the compactified imaginary time direction, Eqs.~\eq{eq_phi_rotation} and \eq{eq_psi_rotation}: the imaginary time translation at the whole period $\beta = 1/T$ is supplemented with a rigid rotation of the three-dimensional space about a fixed spatial axis. The rotation angle~\eq{eq_theta} is determined by the imaginary frequency~$\Omega_I$. Since the rotation takes place in the Euclidean spacetime, it is called the imaginary rotation, as opposed to the real rotation in Minkowski spacetime.

According to the Tolman-Ehrenfest law~\cite{Tolman:1930ona,Tolman:1930zza}, the real rigid rotation of a thermally equilibrated plasma leads to the increase of local temperature towards the outer plasma regions~\eq{eq_T_rho}. We derived the Euclidean version of the TE law~\eq{eq_T_rho_E}, which is valid for imaginary rotations in the imaginary time formalism. The Euclidean version shows the expected periodicity in the imaginary frequency~\eq{eq_periodicity}. It also reduces to the usual TE law after the Wick transformation~\eq{eq_Omega_I_2} to Minkowski spacetime. The TE temperature corresponds to the temperature experienced by a heavy particle (a quark) co-rotating with the thermal ensemble. 

We performed numerical Monte Carlo simulations of $SU(3)$ Yang-Mills theory with the imaginary angular frequency $\theta_I = \pi T/2$. This particular frequency corresponds to the $\pi/2$-rotwisted boundary condition, which is consistent with the symmetries of the hypercubic lattice. The $\pi/2$ rotwisted boundary conditions correspond, using a formal analytical continuation, to a very fast real rotation ($\Omega \sim 300$\,MeV compared to the estimated $\Omega \sim 7$\,MeV in noncentral realistic collisions of heavy ions~\cite{STAR:2017ckg}) at deconfining critical temperature $T_c$.

Using the ``rotwisted'' Polyakov loop defined in the Euclidean co-rotating frame, Fig.~\ref{fig_loops}(c), we found the emergence of the inhomogeneous, spatially non-uniform confining-deconfining phase, which appears due to the imaginary rotation. Our data demonstrates that Euclidean Yang-Mills plasma rotating with the imaginary frequency $\Omega_I = \pi T/2$ possesses the single crossover-type deconfinement transition in the spatial region in the vicinity of the rotating axis at the temperature close to the critical deconfining temperature of the non-rotating gas. While the central near-axis spatial domain experiences the deconfining transition, the outer regions remain in the confining phase.

Our Euclidean results indicate that the gluon plasma, rotating rigidly at real angular frequencies $\Omega$ in Minkowski spacetime, produces a new, spatially non-uniform confining-deconfining phase. This inhomogeneous phase possesses a phase boundary that separates the confining central domain near the rotation axis from the deconfining region in the outer region of the plasma (notice that under the Wick transformation from Euclidean space to Minkowski spacetime, the confining and deconfining regions of the inhomogeneous mixed phase switch their places, Fig.~\ref{fig_phaselines}). This result, which has a pure kinematic origin, confirms the validity of the Tolman-Ehrenfest arguments in the context of the rotating Yang-Mills plasma~\cite{Chernodub:2020qah}.

We also studied the Polyakov loop in the Euclidean laboratory frame. This operator is defined on a contour which winds four times around the compactified time direction, Fig.~\ref{fig_loops}(b). Its expectation value does not show any local sign of the deconfining phase and, surprisingly, no breaking of the global center ${\mathbb Z}_3$ symmetry at a conventional range of lattice couplings $\beta$ that correspond to the deconfining phase transition in the conventional, non-rotating SU(3) Yang-Mills theory. However, the physical interpretation of the {\it global} center ${\mathbb Z}_3$ symmetry can be questioned because our system possesses two {\it locally} separated phases (with broken {\it and} unbroken symmetry, respectively), coexisting in the {\it global} thermal equilibrium.

Our results suggest that we can define two types of temperatures for Euclidean field theory rotating with the imaginary angular frequency. One of them is the local Tolman-Ehrenfest temperature~\eq{eq_T_rho}. The thermal bath with this temperature is experienced by a heavy test particle (or measured by a thermometer) which rotates together with the thermal plasma. The change of the local temperature with rotation is a purely kinematic effect related to a local redshift of the thermal wavelength in curved spacetime. In the context of a confining theory in Euclidean spacetime, the rotating thermal bath temperature is probed by the ``rotwisted'' Polyakov loop, Fig.~\ref{fig_loops}(c). This local Euclidean TE temperature~\eq{eq_T_rho_E} can be analytically continued to the thermal bath temperature~\eq{eq_T_rho} in a rotating system in Minkowski spacetime. 

Another definition of temperature is probed by the Polyakov loop in the laboratory frame, Fig.~\ref{fig_loops}(b). This ``laboratory'' temperature is determined via Eq.~\eq{eq_T_lab}~\cite{Chernodub:2022wsw}. We argued that for $\pi/2 \equiv \frac{1}{4} \times 2\pi$ rotwisted boundary conditions, this laboratory temperature, $T_{\mathrm{lab}} = 1/(4 L_\tau)$, is four times lower than it is naively expected from the length $L_\tau$ of the compactified imaginary-time direction. While this Euclidean laboratory temperature cannot be analytically continued to Minkowski spacetime, it can be associated with heatbath of particles with correct thermal population numbers~\cite{in_preparation}.

The ambiguity in the definition of temperature is a particular feature of the Euclidean imaginary formulation of thermal field theory. The ambiguity does not appear in the Minkowski spacetime where a single equilibrium temperature is identified in a local thermal frame (in our case, the co-rotating frame) using, for example, thermal particle occupation numbers. In the laboratory frame, related to the co-rotating frame by a diffeomorphism transformation (a local Lorentz boost), the particle distribution is not thermal but it is still determined by the temperature in the thermal frame.

The emergence of two notions of temperature becomes apparent after noticing that the order parameter, the Polyakov loop, is a non-local operator which involves a closed path in the imaginary time. For a theory subjected to a rotwisted boundary condition, this non-locality allows us to identify two types of operators, one in the Euclidean rotating frame and another in the Euclidean laboratory frame. The former quantity gives us the free energy of a test quark exposed to the heat bath with a local TE temperature which is related, via a formal analytical continuation, to the real TE temperature in Minkowski spacetime. The meaning of the latter definition of temperature remains to be clarified. 

\begin{acknowledgments}
M.N.C. is grateful to Michele Pepe for discussing the $\pi/2$ imaginary rotation at the Strong and Electro-Weak Matter 2022 conference. 
%(Paris, France, June 20-24, 2022). 
The numerical simulations was performed at the computing cluster Vostok-1 of Far Eastern Federal University. V.A.G. has been supported by RSF (Project No. 21-72-00121). The work of V.A.G. and A.V.M. was supported by Grant No. 0657-2020-0015 of the Ministry of Science and Higher Education of Russia.
\end{acknowledgments}

\end{document}